\documentclass[12pt]{article}
\usepackage{amsmath,amssymb,amsfonts,epsfig,graphicx}
\newcommand{\be}{\begin{equation}}
\newcommand{\ee}{\end{equation}}

\newcommand{\bse}{\begin{subequations}}
\newcommand{\ese}{\end{subequations}}
\newcommand{\bea}{\begin{eqnarray}}
\newcommand{\eea}{\end{eqnarray}}
\newcommand{\ba}{\begin{array}}
\newcommand{\ea}{\end{array}}
\def\L{{\cal L}}

\def\half{\frac{1}{2}}

\def\Ne{${\cal N}=8$\ }
\def\DtNe{$D=3,\ {\cal N}=8\ $}

\def\cK{${\cal K}$}
\def\cG{${\cal G}$}
\def\rta{relaxed-three-algebra}
\def\crta{${\cal RA}_3$}
\def\cKs{${\cal K}_S$}

\def\bbl{\bigl{[}\hspace{-1.2mm}\bigl{[}}
\def\bbr{\bigr{]}\hspace{-1.1mm}\bigr{]}}
\def\one{1\!\!1}
 \makeatletter \@addtoreset{equation}{section}

\makeatother
\begin{document}
\baselineskip 18pt%

\begin{titlepage}
\vspace*{1mm}%
\hfill%
\vbox{
    \halign{#\hfil \cr
\;\;\;\;\;\;\;\;\;\; IPM/P-2008/037 \cr %arXiv:0807.nnnn {\tt
%[hep-th]} \cr
           \cr
           } % end of \halign
      }  % end of \vbox
\vspace*{10mm}%

\centerline{{\Large {\bf  \textsl{The Relaxed
Three-Algebras:}}}} %\centerline{\bf{and}}
\vspace*{3mm}
\centerline{{\large {\bf Their Matrix Representation and Implications for Multi M2-brane Theory}}}%
\vspace*{7mm}
\begin{center}
{\bf \large{M. Ali-Akbari$^{1,2}$, M. M. Sheikh-Jabbari$^{1}$ and J.
Sim\'on$^{3}$  }}
\end{center}
\begin{center}
\vspace*{0.4cm} {\it {$^1$School of Physics, Institute for research
in fundamental
sciences (IPM)\\
P.O.Box 19395-5531, Tehran, IRAN\\
$^2$Department of Physics, Sharif University of Technology\\
P.O.Box 11365-9161, Tehran, IRAN\\
$^3$ School of Mathematics and Maxwell Institute for Mathematical Sciences,\\
King's Buildings, Edinburgh EH9 3JZ, Scotland}}\\
{E-mails: {\tt aliakbari, jabbari @theory.ipm.ac.ir, j.simon@ed.ac.uk}}%
\vspace*{1.5cm}
\end{center}

\begin{center}{\bf Abstract}\end{center}
\begin{quote}
We argue that one can relax the requirements of  the non-associative
three-algebras recently used in constructing $D=3$ \Ne
superconformal field theories, and introduce the notion of ``relaxed
three-algebras''. We present a specific realization of the relaxed
three-algebras in terms of  classical Lie algebras with a matrix
representation, endowed with a non-associative four-bracket
structure which is prescribed to replace the three-brackets of the
three-algebras. We show that both the $so(4)$-based solutions as
well as the cases with non-positive definite metric find a uniform
description in our setting.  We  discuss the implications of our
four-bracket representation for the \DtNe and multi M2-brane theory
and show that our setup can shed light on the problem of negative
kinetic energy degrees of freedom of the Lorentzian case.

\end{quote}
\end{titlepage}
%*******************************************************************************************************************************************
\tableofcontents
\section{Introduction}

Until recently finding an action for the maximally supersymmetric
three-dimensional conformal (gauge) field theory had remained
elusive \cite{BL1,BL2, Gust1, Gust2} (e.g. see \cite{Schwarz-04}
for a short review). The \DtNe superconformal field theory (SCFT) is
expected to arise from  the ``low energy'' effective action describing
many M2-branes on eleven dimensional Minkowski spacetime.
Hence its formulation is closely linked with finding the theory describing
$N$ eleven-dimensional membranes. Furthermore, via the AdS/CFT
correspondence \cite{AdS/CFT}, this SCFT is dual to M-theory on
$AdS_4\times S^7$, the background which is obtained from the
geometry corresponding to coincident parallel M2-branes in the
near-horizon (decoupling) limit \cite{AdS/CFT}.

The \DtNe SCFT action is invariant under the three-dimensional
superconformal group $Osp(8|4)$, with bosonic generators
belonging to $so(8)\times usp(4)\simeq so(8)\times so(3,2)$.
Moreover, the action for a single M2-brane enjoys
invariance under the area preserving diffeomorphisms (APD's) on the
2+1 dimensional world-volume  as its local (gauge) symmetry e.g.
\cite{Hoppe,  Zachos, Isidro}. Thus the multi membrane action is
expected to have a gauge symmetry which somehow manifests this local
gauge invariance. The mathematical (algebraic) structure which
encodes three-dimensional APD's is the Nambu three-bracket
\cite{Isidro, Nambu, Takhtajan}. Therefore, finding an action for
the \DtNe SCFT is ultimately related to quantization of Nambu
three-brackets.

It has been argued that although classical Nambu $p$-brackets
($p\geq 3$) enjoy associativity (e.g. see Appendix B of \cite{TGMT})
the ``quantized'' Nambu $p$-brackets cannot be associative
\cite{Zachos}. For the case of three-brackets, as was proposed
originally by Nambu \cite{Nambu}, one may use the associator of a
non-associative algebra as the quantum version of the three-bracket.
In fact this idea was put at work by Bagger and Lambert to construct
the action for the \DtNe SCFT, the BL theory \cite{BL1, BL2}, where
this non-associative algebra with its three element structure (the
associator) was called the \emph{three-algebra}. A
\emph{three-algebra no-go theorem} was argued for in \cite{Schwarz}
and then proved in \cite{Papadopoulos:2008sk}. This no-go theorem
states that the only three-algebra which has a positive definite
norm is either $so(4)$ or direct sums of a number of $so(4)$'s. In
order to describe $N$ M2-brane theory (for a generic $N$), similarly
to $N$ D$p$-brane cases, one would like to be able to write the BL
theory with more general algebras whose rank (or dimension) are
related to the number of M2-branes and hence bypass this no-go
theorem. This theorem can, however, be circumvented by considering
algebras of non-positive norm \cite{Russo, Verlinde, HIM}.

In this paper we use another prescription for  quantizing the Nambu
three-bracket. This prescription was used in \cite{TGMT} to quantize
type IIB D3-branes to obtain a matrix theory description for the
DLCQ of IIB string theory on the $AdS_5\times S^5$ or the
plane-wave. In this approach we replace the classical Nambu
three-brackets with the ``quantum'' Nambu four-brackets which
involve usual matrices. Although the structure of the quantized
Nambu four-bracket we obtain is non-associative \cite{TGMT} the
underlying algebra, which is nothing but the usual matrix
multiplication algebra, is associative. In particular we use
$2N\times 2N$  matrices to describe the \DtNe SCFT corresponding to
the low energy limit of $N$ M2-branes.

Our prescription requires an extension or relaxation of the notion
of three algebras giving rise to multi M2-brane theories, which will
be called \emph{relaxed three-algebras}. Recently modifications on
the mathematical conditions defining a three-algebra  have been
considered. These ``generalized'' three-algebras are obtained by
relaxing the antisymmetry of the three-bracket and metricity of the
algebra \cite{Gen-BL}. Here, instead of focusing on the antisymmetry
or metricity of these algebras, we will relax the closure and the
fundamental identity conditions in a way to be described below.
 In our representation for the relaxed
three-algebras we show that only the two Euclidean and Lorentzian
cases are possible, compatible with results of
\cite{Papadopoulos:2008sk} and \cite{Jose-Fig}. Moreover, we show
that for the Lorentzian case  the $su(N)$ algebras in $N\times N$
representation are relevant to the theory of $N$ M2-branes. More
importantly we show that there is nothing inherently ``Lorentzian''
in the underlying $su(2N)$ algebra over which the four-bracket
structure is defined.

This paper is organized as follows. In section 2, we give a brief
review of the BL theory and its supersymmetry and gauge
transformations. In section 3, we present the notion of relaxed
three-algebras. In section 4, we derive matrix representations
for the relaxed three-algebras. This is done through the
``four-brackets'' which replace the three-brackets of the BL
three-algebras. We check these representations satisfy the necessary (relaxed)
closure and fundamental identity conditions. In section 5, we
discuss the implications of our relaxed three-algebra realizations
for the multi M2-brane BL theory. We argue that our
prescription, supplemented by arguments of \cite{Sen1, Sen2}, resolves
the problem of ghost-type degrees of freedom appearing in the
ordinary treatment of the Lorentzian case (see \cite{ghost-gauging} for
other ways to resolve the ghost problem). We check that this theory
has the necessary properties expected from a \DtNe SCFT and multi
M2-brane action by examining its behaviour under worldvolume parity
and spectrum of its 1/2 BPS states. The last section is devoted to
concluding remarks and open questions.

\section{Review of the BLG theory}
In this section to fix the conventions and notations we briefly
review the BLG theory by first defining the three-algebras ${\cal
A}_3$ and their algebraic structure and then presenting the BLG
proposed action for the \DtNe superconformal field theory.

\subsection{The BLG three-algebras}\label{BLG-three-algebras-section}

The three-algebra ${\cal A}_3$ is an algebraic structure defined
through the three-bracket $\bbl\ , \ ,\ \bbr$%
\be\label{3-bracket-def}%
\bbl \Phi_1, \Phi_2, \Phi_3\bbr \in {\cal A}_3, \quad {\rm{for\
any}\ } \Phi_i\in {\cal A}_3, %
\ee%
where%
\be\label{total-anti-smmetry}%
\bbl \Phi_1,\Phi_2,\Phi_3\bbr =- \bbl \Phi_2,\Phi_1,\Phi_3\bbr =-
\bbl \Phi_1,\Phi_3,\Phi_2\bbr%
\ee%
 The three-bracket, which is a ``quantized'' Nambu three-bracket \cite{Nambu}
is indeed an associator and  ${\cal A}_3$ is a
\emph{non-associative} algebra. The three-bracket should satisfy
an analog of the Jacobi identity, the \emph{fundamental identity} \cite{Takhtajan}:%
\be\label{fund-ident}\begin{split}%
{\cal K}_{ijklm}&={\bbl} \bbl \Phi_i,\Phi_j,\Phi_k\bbr, \Phi_l,
\Phi_m{\bbr}+ {\bbl} \bbl \Phi_i,\Phi_j,\Phi_l\bbr, \Phi_m,
\Phi_k{\bbr}+{\bbl} \bbl \Phi_i,\Phi_j,\Phi_m\bbr,\Phi_k,
\Phi_l{\bbr} \cr  &= {\bbl}
\Phi_i,\Phi_j,\bbl \Phi_k, \Phi_l, \Phi_m\bbr{\bbr}.%
\end{split}
\ee%
As we can see ${\cal K}_{ijklm}$ is anti-symmetric under exchange of the
first two  as well as the last three indices.
We equip this algebra with a product $\bullet$ and a Trace%
 \be\label{trace-def}%
 Tr(\Phi_1\bullet \Phi_2)= Tr (\Phi_2\bullet \Phi_1) \in\mathbb{C}
\ee%
satisfying a ``by-part integration'' property%
\be\label{push-thru-trace}%
 Tr(\Phi_1\bullet \bbl\Phi_2, \Phi_3, \Phi_4\bbr )=-
Tr(\bbl\Phi_1,\Phi_2, \Phi_3\bbr \bullet \Phi_4).%
\ee%

For the usage in physical theories, noting that $\Phi_i$'s are
complex valued, it is needed to define the Hermitian conjugation
over the algebra. In particular if we choose to work with Hermitian
algebras, i.e.%
\[ \Phi^\dagger=\Phi, \quad \forall \Phi\in {\cal A}_3, \]%
then the closure condition \eqref{3-bracket-def} is satisfied with
the following definition for complex conjugation of the
three-bracket:%
\be\label{bracket-dagger}%
 \bbl \Phi_1, \Phi_2,\Phi_3\bbr^\dagger
=\bbl \Phi_1^\dagger, \Phi_2^\dagger, \Phi_3^\dagger\bbr\ .%
\ee%

If we expand  ${\cal A}_3$ elements in terms of the complete
basis $T^a$ \[ \Phi=\Phi_a T^a \] then \eqref{3-bracket-def} implies that%
\be\label{structure-const}%
\bbl T^a, T^b,T^c\bbr = f^{abc}_{\quad\ d} T^d
\ee%
and%
\be\label{metric-def}%
Tr (T^a\bullet T^b)\equiv h^{ab}
\ee%
defines the metric $h^{ab}$ on  ${\cal A}_3$. Mathematically, the
metric $h^{ab}$ can have arbitrary signature, though physically,
non-positively defined signatures could give rise to ghost degrees
of freedom. We will always take $h^{ab}$ to be non-degenerate and
invertible.  Noting  \eqref{push-thru-trace},
\[
f^{abcd}\equiv f^{abc}_{\quad\ e} h^{ed},
\]
is totally anti-symmetric four-index structure constant. The
fundamental identity in terms of the structure constant $f$ is
written as%
\be\label{fund-iden-f}%
f^{abc}_{\quad\ l}f^{del}_{\quad\ m}+f^{abd}_{\quad\
l}f^{ecl}_{\quad\ m}+f^{abe}_{\quad\ l}f^{cdl}_{\quad\ m}
=f^{cde}_{\quad\ l}f^{abl}_{\quad\ m}.%
\ee%
This equation does not have any solution other than
$f^{abcd}=\epsilon^{abcd}$ or four tensors made out of
$\epsilon^{abcd}$,  if $h^{ab}$ is positive definite and hence
${\cal A}_3$ is either $so(4)$ or combinations involving the direct
sums of $so(4)$ \cite{Papadopoulos:2008sk}.

To find three-algebras other than $so(4)$ one is hence forced to
relax the positive definite condition on $h_{ab}$
\cite{Russo,Verlinde}. Explicitly if we choose $a=(+,-,\alpha)$
and %
\be\label{metric-signature}%
h_{\alpha\beta}=\delta_{\alpha\beta},\
h_{+\alpha}=h_{-\alpha}=0,\ h_{++}=h_{--}=0,\ h_{+-}=h_{-+}=-1%
\ee%
then $f^{abc}_{\quad\ d}$ with non-zero components%
\be\label{f-verlinde}%
f^{\alpha\beta\gamma}_{\quad\ -}\equiv f^{\alpha\beta\gamma},\
f^{\alpha\beta +}_{\quad\ \gamma}=f^{\alpha+\beta}_{\quad\
\gamma}=-f^{\alpha+\beta }_{\quad\ \gamma}=
f^{\alpha\beta\rho}\delta_{\rho\gamma}%
\ee%
is a solution to the fundamental identity \eqref{fund-iden-f},
provided that $f_{\alpha\beta\gamma}$ are satisfying the usual
Jacobi identity for associative algebras \cite{Verlinde}.

Finally we point out that if $T^a$'s are all Hermitian then with
\eqref{bracket-dagger} the structure constant $f_{abcd}$ should be
real valued, that is $f_{abcd}^*=f_{abcd}$.

\subsection{The BLG action}

The  on-shell matter content of the \DtNe hypermultiplet involves
eight three-dimensional scalars $X^I$, $I=1,2,\cdots ,8$ in the
$8_v$ of the $SO(8)$ R-symmetry group, eight two component
three-dimensional fermions $\Psi$ in the $8_s$  of SO(8) (we have
suppressed both the $3d$ and the R-symmetry  fermionic indices).
Each of the above physical fields which will generically be denoted
by $\Phi$ are also assumed to be elements of the three-algebra and
hence
\[\Phi= \Phi_a T^a.\]%

The action of the BLG \cite{BL1, BL2, Gust1} theory is given by%
\be\label{BL-action} \begin{split}%
  S&=\int d^3\sigma\ Tr\bigg(-\half D_i
  X^ID^i X^I-\frac{1}{2.3!}\bbl X^I,X^J,X^K\bbr \bbl
  X^I,X^J,X^K\bbr \cr &\qquad +\frac{i}{2}\bar{\Psi}\gamma^i
  D_i\Psi-\frac{i}{4}\bbl \bar{\Psi},X^I,X^J\bbr \Gamma^{IJ}\Psi\biggr)%
   +\L_{twisted\ Chern-Simons}
\end{split}\ee %
where $\L_{twisted\ Chern-Simons}$ is a parity invariant Chern-Simons action%
\be\label{twisted-CS}%
\L_{twisted\ Chern-Simons} =\frac{1}{2}\epsilon^{ijk}\left(f^{abcd}
A_{i\ ab}\partial_j A_{k\ cd}+\frac{2}{3} f^{abcl}f^{deg}_{\quad\ l}
A_{i\ ab} A_{j\ de} A_{k\ cg}\right).%
\ee%
Indices $i=0,1,2$ denote the three-dimensional directions and the
covariant derivatives are defined as%
\be\label{cov-der}%
(D_i \Phi)_a \equiv \partial_i \Phi_a - f^{cdb}_{\ \ \ \ a}\ {A}_{i\ cd} \Phi_b %
\ee%
where ${A}_{i\ ab}$ is the non-propagating three dimensional,
two-index gauge field.  For later use it is useful to introduce
another gauge field
\be\label{Avs.tildeA}%
{\tilde A}_{i\ \ a}^{\ b}=f^{cdb}_{\ \ \ \ a}\ {A}_{i\ cd}.%
\ee%

The above action is invariant under the local gauge symmetry:
\be\label{gauge-sym}\begin{split} %
\delta_{gauge} \Phi_a &= f^{cdb}_{\ \ \ \ a}{\Lambda}_{cd}\Phi_b, \cr%
\delta_{gauge} A_{i\ cd} &= \partial_i \Lambda_{cd}-f^{abe}{}_{[c}\Lambda_{d ]e} A_{i\ ab}   %
\end{split}
\ee%
as well as the global supersymmetry transformations%
\be\label{SUSY-trans}\begin{split}%
\delta_{susy} X^I &= i\bar{\epsilon}\Gamma^I\Psi \cr %
\delta_{susy}\Psi &=D_i X^I\Gamma^I\gamma^i\epsilon-\frac{1}{6}\bbl X^I,X^J,X^K \bbr \Gamma^{IJK}\epsilon \cr%
\delta_{susy} \tilde{A}_{i}^{ab} &=if^{abcd} \epsilon\gamma_i
\Gamma_{I} X^I_{c}\Psi_d
\end{split}\ee %
It has also been shown that \cite{Schwarz} besides the $2+1$
dimensional super-Poincar\'e symmetry the above action, at least at
classical level, is invariant under the full three-dimensional
superconformal algebra.

The equations of motion of the above action are
\be\label{e.o.m}\begin{split} %
 &\gamma^i D_i\Psi+\half\Gamma^{IJ}\bbl X^I,X^J,\Psi\bbr =0\cr
 &D^2X^I-\frac{i}{2}\Gamma^{IJ}\bbl \bar{\Psi},X^J,\Psi\bbr+\half \bbl X^J,X^K,\bbl X^I,X^J,X^K\bbr\bbr=0\cr
 &\tilde{F}_{ij}^{ab}+\epsilon_{ijk} f^{abcd}\big( X^J_c D^k X^J_d+\frac{i}{2}\bar{\Psi}_c\gamma^k\Psi_d\big)=0
\end{split}\ee %
where
\[
\tilde{F}_{ij}^{\ \  b}{}_{\ a}=\partial_{i} \tilde{A}_{j}^{\
b}{}_{\ a}-\partial_{j} \tilde{A}_{i}^{\ b}{}_{\ a}-
\tilde{A}_{i}^{\ b}{}_{\ c}\tilde{A}_{j}^{\ c}{}_{\
a}+\tilde{A}_{j}^{\ b}{}_{\ c}\tilde{A}_{i}^{\ c}{}_{\ a} \ .
\]
 In the BL theory, for both the Lorentzian and Euclidean
realizations of the three-algebras, the basis $T^a$ and hence all
the components of the $X$ field $X_a$ are both taken to be
Hermitian. It is also worth noting that with this requirement and
the Hermiticity property \eqref{bracket-dagger} the \emph{potential}
terms in the Hamiltonian of the BL theory in both Lorentzian and
Euclidean cases are positive definite.

\section{The Relaxed Three-Algebras}\label{relaxed-section}

The construction of BLG three-algebras with the definition and
properties outlined in section \ref{BLG-three-algebras-section} has
proven very restrictive. In this section we revisit the BL analysis
with the idea that we may be able to relax some of the conditions on
the BL three-algebras while keeping the physical outcomes intact. As
we will show this is indeed possible.

As discussed  in section \ref{BLG-three-algebras-section},
three-algebras of interest are defined by five conditions: a totally
anti-symmetric three-bracket, existence of non-degenerate metric,
the closure of the three-algebra under the three-bracket, the
fundamental identity and the trace property \eqref{push-thru-trace}.
The antisymmetry, closure and fundamental identity are conveniently
expressed in terms of a basis $T^a$ and the structure constants
$f^{abc}{}_{d}$ as in \eqref{structure-const} and
\eqref{fund-iden-f}.

Let us \emph{relax} the closure and fundamental identities, while
keeping the antisymmetry and the trace property, by enlarging the
set of $T^a$'s through the addition of
extra generators $T^A$'s satisfying the properties:\\
\emph{i) $T^A$ is orthogonal to every other generator, i.e.}%
\be\label{Orthogonality}%
Tr(T^a T^A)=0, \qquad Tr(T^A T^B)=0\ . %
\ee%
\emph{ii) $T^A$ in the brackets:}%
\bse\label{TA-three-brackets}%
\begin{align}%
&\bbl T^a, T^b, T^c \bbr = f^{abc}{}_d T^d+ k^{abc}{}_{A} T^A,\\
\bbl T^a, T^b, T^A \bbr = f^{abA}{}_B T^B , \ &\bbl T^a, T^A, T^B
\bbr = f^{aAB}{}_C T^C,  \ \bbl T^A, T^B, T^C\bbr = f^{ABC}{}_D T^D,
\end{align}%
\ese%
where $f^{abc}{}_d$ are still satisfying the standard fundamental
identity \eqref{fund-iden-f} and any other additional four-index
structure constant, i.e. $f^{xyz}{}_A$ $\forall\,\,x,\,y,\,z$, is
yet unknown. Notice that the form of $\bbl T^a, T^b, T^A\bbr$,
$\bbl\ T^a, T^A, T^B\bbr$ and $\bbl T^A, T^B, T^C\bbr$ is fixed by
demanding the consistency of these brackets with the ``by-part''
property \eqref{push-thru-trace}. In this sense
\eqref{TA-three-brackets} is a consequence of \eqref{Orthogonality}
and not an independent assumption.

If $k^{abc}{}_A$ are zero, we can just simply ignore the
existence of the $T^A$'s and we are back to the BL three-algebra
${\cal A}_3$. However, with non-zero $k^{abc}{}_A$, the algebra of
$T^a$'s does not close. It is evident that even if we ignore the
non-closure of the three-algebra based on $T^a$'s the fundamental
identity for the extended algebra does not hold. Nonetheless, we can
still have a generalized or \emph{relaxed} notion of closure. If we
denote the part of the algebra spanned by $T^a$'s by \cK\ and the
part spanned by $T^A$'s by \cKs , \eqref{TA-three-brackets} can be
rewritten as
\bse\label{relaxed-closure}%
\begin{align}
\bbl \Phi_1,\Phi_2,\Phi_3 \bbr &\in {\cal K}\oplus {\cal K}_S,
\qquad \forall \Phi_i \in {\cal K}\, ,\\
\bbl \Phi_1,\Phi_2,\chi \bbr &,\ \bbl \Phi,\chi_1,\chi_2 \bbr ,\
\bbl \chi_1,\chi_2, \chi_3 \bbr\in {\cal K}_S \qquad \forall\
\Phi_i\in {\cal K},\ \chi_i\in {\cal K}_S \ .
\end{align}
\ese%
Therefore, with the above it is immediate to see that if we shift an
element of \cK\ by an arbitrary element in \cKs, the part of the
resulting bracket which resides in \cK\ does not change. In this
sense \eqref{relaxed-closure} defines the notion of \emph{relaxed
closure} over \cK.

It will be convenient to introduce the notion of ``physical'' part
of a given three-bracket. Let $\Upsilon_i$ be a general element in
\cK$\ \oplus\ $\cKs. It can then be decomposed into
its \emph{physical} part $\Phi_i$ (which is in \cK ) and its
\emph{spurious} part
$\chi_i$ (which is in \cKs ). In other words,%
\be\label{Upsilon-physical}%
(\Upsilon_i)_{phys}=\Phi_i= h_{ab} T^a\ Tr(T^b\Upsilon_i), \quad
\forall
\Upsilon_i\in {\cal K}\oplus {\cal K}_S\ , %
\ee%
where $h_{ab}$ is the inverse of the metric $h^{ab}=Tr(T^a T^b)$. It is
also useful to note that%
\be%
Tr(\Phi \chi)=0 \ , \qquad \forall \Phi\in {\cal K},\ \chi\in {\cal
K}_S\ ,%
\ee%
and,%
\be\label{physical-bracket}%
\left(\bbl \Upsilon_1, \Upsilon_2, \Upsilon_3\bbr \right)_{phys}=
\left(\bbl \Phi_1, \Phi_2, \Phi_3\bbr\right)_{phys}=f^{abc}{}_d\
\Phi_{1a}\Phi_{2b}\Phi_{3c} T^d\ . %
\ee%
In terms of the physical part of a bracket, the relaxed closure condition is
nothing but the closure for the physical part of the brackets.

In the same spirit as above one may define a notion of
\emph{relaxed} fundamental identity, by demanding the fundamental
identity \eqref{fund-ident} to hold for the physical
part of the three brackets. Explicitly,%
\be\label{relaxed-fund-id}%
\begin{split}%
{\bbl} \bbl \Upsilon_i,\Upsilon_j,\Upsilon_k\bbr_{phys}, \Upsilon_l,
\Upsilon_m{\bbr}_{phys} + {\bbl} \bbl
\Upsilon_i,\Upsilon_j,\Upsilon_l&\bbr_{phys}, \Upsilon_m,
\Upsilon_k{\bbr}_{phys} + \cr{\bbl} \bbl
\Upsilon_i,\Upsilon_j,\Upsilon_m\bbr_{phys},\Upsilon_k,
\Upsilon_l{\bbr}_{phys}  &= {\bbl}
\Upsilon_i,\Upsilon_j,\bbl \Upsilon_k, \Upsilon_l, \Upsilon_m\bbr_{phys}{\bbr}_{phys}.%
\end{split}
\ee%
In terms of the structure constants $f$, this is equivalent to
requiring $f^{abc}{}_d$ to satisfy \eqref{fund-iden-f}.

With above notion of the relaxed closure and fundamental identity,
together with the orthogonality properties \eqref{Orthogonality}, we
define a \emph{relaxed-three-algebra} (${\cal RA}_3$). Any given
\crta\ has a \emph{physical} part \cK\ and an \emph{spurious} part
\cKs .

Let us now rewrite the BLG theory with the above
\emph{relaxed-three-algebra} by adding $T^A$ components to the
physical fields, i.e. we take the  fields to be%
\be\label{exntended-Phi}%
\Upsilon=\Phi_a T^a+ \chi_A T^A\ ,%
\ee%
and let the gauge fields to also have $A_{i\ aA}$ components. With
the trace conditions \eqref{Orthogonality} it is readily seen that
the $\chi_A$ components of the fields do not appear in the action at
all.  This is very similar to the notion of physical and spurious
states in a $2d$ CFT e.g. see \cite{Polch1}. Since the action does
not involve the spurious fields the equations of motion for the
physical fields will not change compared to the ordinary BL case.

One can also check the supersymmetry and the gauge symmetry
invariance of the action within the relaxed-three-algebra. The
only part which should be checked is where the fundamental
identity is used. As discussed in \cite{BL2}  the fundamental
identity is needed for the closure of supersymmetry when two
successive supersymmetry transformations on the gauge field is
considered. One can, however, see that with the structure of the
three-brackets introduced in \eqref{TA-three-brackets}, the part
in equation (35) of \cite{BL2} does not harm the closure of the
supersymmetry algebra as long as $f^{abc}{}_d$ are still
satisfying the fundamental identity \eqref{fund-iden-f}.

In the next section we will give a  construction based on usual
matrices which realizes this relaxed-three-algebras \crta .

\section{Matrix representation for the relaxed-three-algebras}

There are many three-algebras, already among the ones having a
bi-invariant metric with Euclidean and Lorentzian signatures, and
one may wonder whether by introducing some additional structure in
the theory reviewed above, one may get stronger constraints on the
classical Lie algebras underlying them. Inspired by the ideas of
\cite{TGMT} regarding quantization of Nambu three-brackets using
four-brackets, we propose to realize the three-bracket in terms of a
four-bracket: \footnote{For
other attempts to find matrix representations for quantized Nambu brackets see \cite{Corneliu}.} %
\be\label{threeVsfour}%
\bbl A, B, C\bbr \equiv [ \hat A, \hat B, \hat C, T^-] %
\ee%
where the hatted quantities are just normal matrices, $T^-$ being among them
(to be specified shortly) and the four-bracket is defined as%
\be\label{four-bracket-def}\begin{split}%
[\hat A_1, \hat A_2, \hat A_3, \hat A_4] &=\frac{1}{4!}
\epsilon^{ijkl} \hat A_i \hat A_j \hat A_k \hat A_l \cr
 &=\frac{1}{4!}\left(\{[\hat A_1, \hat A_2], [\hat A_3, \hat A_4]\}-\{[\hat A_1, \hat A_3], [\hat A_2, \hat
 A_4]\}+\{[\hat A_1, \hat A_4], [\hat A_2, \hat A_3]\}\right).
\end{split} %
\ee%
The fundamental identity \eqref{fund-ident} in terms of the
four-bracket takes the form\footnote{For the ease of notation, we will omit the
hats $\hat{A}$ on any matrix $A$. It should be clear from the bracket under consideration
the nature of the object under consideration.}
\be\label{fund-iden-4bracket}\begin{split}%
[[A, B,C, T^-], D, E ,T^-]&+[C,[A,B,
D, T^-],E, T^-] \cr &+[C,D,[A,B,E,T^-],T^-]=[A,B, [C,D,E, T^-],T^-]. %
\end{split}\ee%

It is straightforward to see that the above four-bracket defines a
non-associative structure over the algebra of matrices and the Trace
over the matrices is the natural trace operation over this algebra.
The Hermitian conjugation of the underlying algebra structure
naturally extends to the four-bracket. If $T^-$ is Hermitian it is
immediate to see that \eqref{bracket-dagger} holds. As we will show
for one of the only two possibilities for $T^-$, $T^-$ is Hermitian.

In the rest of this section we show that the above proposal
\eqref{threeVsfour}, within the setup of the
\emph{relaxed-three-algebras} of previous section, works for the two
currently recognized three-algebras, namely the $so(4)$-based
algebras \cite{Papadopoulos:2008sk} and those coming with a
Lorentzian signature metrics of \cite{Russo,Verlinde,Jose-Fig}. In
fact, within our working assumptions described below,
these are the only two possible cases.

 From the definition  it is directly seen that the four-bracket has
the anti-symmetry property \eqref{total-anti-smmetry}. Using the
explicit definition \eqref{four-bracket-def} and standard matrix
algebra, it is straightforward to see that the ``by-part
integration'' property \eqref{push-thru-trace} is also satisfied. We
are then left with verifying the (relaxed) closure and fundamental
identities.
%However, as discussed for the purpose of using the BLG
%theory we only need the ``relaxed'' closure and fundamental
%identities.

\subsection{The relaxed closure and fundamental identities}

All the elements we consider belong to a finite dimensional matrix
representation of an underlying Lie-algebra \cG. \cG\ is an ordinary
(classical) Lie-algebra defined through commutator relations and
ordinary structure constants. The three-bracket structure is,
however, defined over a subset of \cG. This subset has two parts:
\cK\ with the basis $T^a$, and \cKs\ with the basis $T^A$. \cK\
contains the  ``physical fields'' and \cKs\ the ``spurious fields"
(\emph{cf.} discussions of section 3). We should emphasize that,
although both \cK\ and \cKs\ are subsets of \cG\ they are not
necessarily sub-algebras of \cG.

The \emph{relaxed} closure conditions \eqref{relaxed-closure} for
three-algebras within our four-bracket structure are then written as
\bse\label{relaxed-closure-four-bracket}%
\begin{align}
[\Phi_1,\Phi_2,\Phi_3, T^-] &\in {\cal K}\oplus {\cal K}_S, \qquad
\forall \Phi_i \in {\cal K}\, ,\\
[\Phi_1,\Phi_2,\chi, T^-], \ [\Phi_1,\chi_1,\chi_2, T^-],&\
[\chi_1,\chi_2, \chi_3, T^-]\in {\cal K}_S \qquad \forall \Phi_i\in
{\cal K},\ \chi_i\in {\cal K}_S \ .
\end{align}
\ese%
In fact we can view the above closure conditions as the definitions
for the subsets \cK\ and \cKs\ in \cG.

For the relaxed-three-algebras \crta\ we demand a
relaxed version of the fundamental identity \eqref{relaxed-fund-id}.
Namely, we only demand \emph{the non-spurious part of the brackets in
(\ref{relaxed-closure-four-bracket}a) to satisfy the fundamental
identity.}

Since $T^-$ has a distinct role in our four-bracket construction, we must
specify it separately. From the closure
conditions \eqref{relaxed-closure-four-bracket} and the definition of
the four-bracket \eqref{four-bracket-def} it is evident that $T^-$
is either in \cK\ or \cKs. To obtain a non-trivial interacting theory,
$T^-$ cannot be in \cKs. This can be seen by recalling the trace conditions
\eqref{Orthogonality} on the spurious parts. Thus we take $T^-$ to be in \cK.

To proceed we will choose  $T^-$ to be an element of \cK, such that its
\emph{anti-commutator} with any element of
\cK\ and \cKs\ is in the center of the underlying algebra \cG,
as our \emph{working assumptions}. In terms of the basis $T^a$ and
$T^A$ this means that either $T^-$
anticommutes with $T^a$ and $T^A$, or its anti-commutator with them
is the identity matrix:
\be\label{anti-commute-T-}%
\begin{split}
 \{ T^-, T^a\}= 0, \quad  {\rm or}&  \quad   \{T^-, T^a\}=\one.\cr %
\{T^-, T^A\} &=0.%
\end{split}%
\ee%
(Note that $\{T^-, T^A\}=\one$ case is not possible due to the trace
condition \eqref{Orthogonality}.)

With the above choice it is evident that any linear
combination of a given set of $T^a$'s is also satisfying the above
anti-commutator conditions. Therefore, within the set of $T^a$'s one
can identify a \emph{single} element whose anti-commutator with
$T^-$ is the identity matrix. We will denote this element by $T^+$.
As $T^- \in {\cal K}$, $T^-$ should then square to zero or to $(1/
2) \one$. Hence, given our working assumptions, there are two cases
to consider for our four-bracket realization of the three-bracket:
\begin{itemize}
\item[i)] $T^-=T^+$, corresponding to $2(T^-)^2=\one$.
\item[ii)] $(T^-)^2=0$, corresponding to $T^+\neq T^-$ and $\{T^+, T^-\}=\one$.

\end{itemize}

We will denote the elements in ${\cal K}$ by $\{T^a\} = \{T^+,\ T^-, T^\alpha\}$.
Without loss of generality, one can always choose the basis such that%
\be\label{basis-choice}%
\{T^{\pm}, T^\alpha\}=0, \qquad \{T^+, T^-\}=\one\ .%
\ee%
Note that while $\{T^+, T^A\}$ can be non-vanishing, it is always
traceless (\emph{cf.} \eqref{Orthogonality}).

We will choose the $T^\alpha$ matrices to be hermitian,
\be\label{hermiticity-Talpha}%
(T^\alpha)^\dagger=T^\alpha \ , %
\ee%
therefore, the metric%
\[h^{\alpha\beta}=Tr(T^\alpha T^\beta),\]%
is positive definite.
Recalling
\eqref{basis-choice},%
\be\label{h+-alpha}%
h^{\pm\alpha}=Tr (T^{\pm} T^\alpha)=0.
\ee%

Thus, the  $T^-=T^+$ case corresponds to a positive definite metric
$h^{ab}$ since $2(T^-)^2=\one$, and consequently $h^{--}$ is
positive. For this case $T^-$ is Hermitian. On the other hand, the
$T^-\neq T^+$ case has Lorentzian signature. This is because $\{T^-,
T^+\}=\one$, and so $h^{-+}=h^{+-}$ is positive definite and
$h^{--}=0$. Hence
\be\label{det-h}%
\det h_{ab}=-\det h_{\alpha\beta}\cdot (h^{+-})^2 < 0.%
\ee%
As $(T^-)^2=0$, for the Lorentzian case $T^-$ cannot be Hermitian.
One can always find a linear combination of $T^+$ and $T^-$ for
which both $h^{--}$ and $h^{++}$ vanish. Here we choose to work in
such a basis.

Equipped with the above we are now ready to examine the relaxed
closure condition \eqref{relaxed-closure-four-bracket} and the
fundamental identity  and check which algebras are satisfying the
above requirements.

%\subsubsection{The $T^-=T^+$ case}
\subsubsection{The Euclidean signature case}

For this case the only non-vanishing ``physical'' four-bracket is of
the form $[T^\alpha, T^\beta, T^\gamma, T^-]$ which recalling
\eqref{anti-commute-T-} can be written as%
\be\label{four-bracket-non-zero}%
[T^\alpha, T^\beta, T^\gamma, T^-]= F^{\alpha\beta\gamma} T^-\,,
\ee%
where $F^{\alpha\beta\gamma}$ is the totally anti-symmetric three-form
\be\label{F-tensor}%
F^{\alpha\beta\gamma}=\frac{1}{12}\left(\{T^\alpha, [T^\beta,
T^\gamma]\}+ \{T^\gamma, [T^\alpha, T^\beta]\}+ \{T^\beta,
[T^\gamma, T^\alpha]\}\right).
%-T^\beta T^\alpha T^\gamma -T^\gamma T^\beta T^\alpha - T^\alpha T^\gamma T^\beta
%\right).%
\ee%
Note that by definition $F^{\alpha\beta\gamma}$ is not necessarily
in the algebra \cG, but in general in its enveloping algebra.

The relaxed closure condition \eqref{relaxed-closure-four-bracket}
demands
$F^{\alpha\beta\gamma}T^- \in {\cal K}\oplus {\cal K}_S$. Equivalently, %
\be\label{F3T-}%
[T^\alpha, T^\beta, T^\gamma, T^-]=f^{\alpha\beta\gamma}_{\quad\
\lambda} T^{\lambda}+ g^{\alpha\beta\gamma} T^-+ k^{\alpha\beta\gamma}{}_A T^A%
\ee%
where $f,\ g$ and $k$ are expansion coefficients, anti-symmetric in
$\alpha\beta\gamma$ indices.

Multiplying both sides of \eqref{F3T-} with $T^-$ and taking trace
of both sides implies that $g^{\alpha\beta\gamma}=0$ and hence we
only remain with $f^{\alpha\beta\gamma}_{\quad\ \lambda}$ and
$k^{\alpha\beta\gamma}{}_A$ terms.

The relaxed fundamental identity \eqref{relaxed-fund-id} then
requires:
\be\label{f-g-eqn}%
%\begin{align}%
f^{\alpha\beta\gamma}_{\quad\ \sigma}\ f^{\sigma\rho\lambda}_{\quad\
\delta}+ f^{\alpha\beta\rho}_{\quad\ \sigma}\
f^{\gamma\sigma\lambda\delta}+ f^{\alpha\beta\lambda}_{\quad\
\sigma}\ f^{\gamma\rho\sigma}_{\quad\ \delta}=
f^{\gamma\rho\lambda}_{\quad\ \sigma}\
f^{\alpha\beta\sigma}_{\quad\ \delta},
%\\%
%f^{\alpha\beta\gamma}_{\quad\
%\sigma}\ g^{\sigma\rho\lambda}+ f^{\alpha\beta\rho}_{\quad\
%\sigma}\ g^{\gamma\sigma\lambda}+ f^{\alpha\beta\lambda}_{\quad\
%\sigma}\ g^{\gamma\rho\sigma}&=
%f^{\gamma\rho\lambda}_{\quad\ \sigma}\ g^{\alpha\beta\sigma}.%
%\end{align}%
\ee%

Since $h^{ab}$ is positive definite, it was proved in \cite{Papadopoulos:2008sk}
that the  \emph{unique} solution to \eqref{f-g-eqn} is given by%
\be\label{f-solution}%
f^{\alpha\beta\gamma\rho}= \epsilon^{\alpha\beta\gamma\rho}%
\ee%
and $\alpha,\beta,\gamma, \rho=1,2,3,4$.  The explicit solution for
this case, as has been discussed in \cite{TGMT, TGMT-1/2-BPS} is
\be\label{T+=T-case}%
T^\alpha={\cal J}^\alpha, \qquad T^-={\cal L}_5%
\ee%
where ${\cal J}^\alpha$ and ${\cal L}_5$  are in general $2J\times
2J$ representation of $so(4)$, which are generalization of the
ordinary $SO(4)$ Dirac gamma matrices
\cite{TGMT-1/2-BPS}.\footnote{As mentioned in
\cite{Papadopoulos:2008sk} direct sums of an arbitrary $so(4)$
algebras also leads to $f^{\alpha\beta\gamma\rho}=
\epsilon^{\alpha\beta\gamma\rho}$.} For $J=2$ they reduce to
$\gamma^\alpha$ and $\gamma^5$. (For an explicit matrix form and
more detailed discussion see \cite{TGMT,TGMT-1/2-BPS}.) Note that
the size of the representation is not fixed by the above
considerations.

The above explicit representation for $T^\alpha$'s leads to
$k^{\alpha\beta\gamma}{}_A=0$ and hence for this case, the Euclidean
case, the \crta\ is the same as the corresponding ordinary BL
three-algebra.

%For \eqref{f-solution} the equation for $g^{\alpha\beta\gamma}$,
%(\ref{f-g-eqn}b) gives $g^{\alpha\beta\gamma}=0$ (recall that
%$SO(4)$ does not have invariant three-forms).

In summary, our four-bracket representation for the three-algebra
and its three-bracket has all the needed properties of the
three-bracket and  the only solution to this case is the
$SO(4)$-based solutions discussed in \cite{Papadopoulos:2008sk}.

Finally it is notable that in this case the algebra \cG\ which is
the algebra generated from ${\cal J}^\alpha$ and ${\cal L}_5$ (and
their commutators) is $so(6)\simeq su(4)$. Note, however, that the
${\cal J}^{5\alpha}=i[{\cal J}^\alpha, {\cal L}_5]$ are not the
$T^A$'s, as they do not satisfy the trace condition
\eqref{Orthogonality} and (\ref{TA-three-brackets}b).

%\subsubsection{ The $(T^-)^2=0$ case}
\subsubsection{The Lorentzian signature case}

There are two different non-vanishing  four-brackets of
``physical'' elements to consider:
\bse\label{T-2=0-four-bracket}%
\begin{align}%
[T^\alpha, T^\beta, T^\gamma, T^-]&= F^{\alpha\beta\gamma} T^-\\
[T^\alpha, T^\beta, T^+, T^-] &= \frac14\,[T^\alpha, T^\beta] T \ ,
\end{align}%
\ese%
where $F^{\alpha\beta\gamma}$ is defined in \eqref{F-tensor} and
\be\label{T-def}%
T\equiv [T^+,T^-]. \ee %
In deriving these expressions, we have used the fact that, by
definition, $T$ commutes with $T^\alpha$ $([T,\,T^\alpha]=0)$.
Furthermore, from $(T^-)^2=0$ and $\{T^+, T^-\}=\one$, one has%
\bse\label{T2}%
\begin{align}%
T^2&=-(\one-2T^+T^-)(\one-2T^-T^+)=\one, \\ %
T T^-&=-(\one-2T^+T^-)T^-=-T^-\ .%
\end{align}%
\ese%

Let us analyze the (relaxed) closure conditions. First, requiring
$[T^\alpha, T^\beta, T^\gamma,T^-]\in {\cal K} \oplus {\cal K}_S$ implies that in the most general form%
\be\label{closure-F3T-}%
F^{\alpha\beta\gamma}T^-=f^{\alpha\beta\gamma}_{\quad\
\lambda}T^\lambda+ g^{\alpha\beta\gamma} T^-+
l^{\alpha\beta\gamma}T^+ + k^{\alpha\beta\gamma}{}_A T^A %
\ee%
where $f$, $g$, $k$ and $l$ are some unknown arbitrary expansion
parameters which are totally anti-symmetric under exchange of
$\alpha$, $\beta$ and $\gamma$ indices. Multiplying both sides of
\eqref{closure-F3T-} in $T^-$ and taking the trace, noting that the
left-hand-side vanishes identically, we learn that
$l^{\alpha\beta\gamma}=0$. Noting that $(T^-)^2=0$ and $\{ T^-, F^{\alpha\beta\gamma}\}=0$ then%
\[
%be\label{useful-anti-com}%
\{T^-, F^{\alpha\beta\gamma}T^-\}=0,\qquad [T^-, F^{\alpha\beta\gamma}T^-]=0 ,%
\]%ee%
and therefore \footnote{Here we will assume working with the
non-trivial case of  $T^\alpha T^-\neq 0$.}
\be\label{f-k}%
T^-\left(f^{\alpha\beta\gamma}{}_{\lambda} T^\lambda+k^{\alpha\beta\gamma}{}_A  T^A\right)=0 . %
\ee%

The second relaxed closure requirement, $[T^\alpha, T^\beta, T^+,
T^-]\in
{\cal K} \oplus {\cal K}_S$ implies,%
\be\label{T+--closure}%
[T^\alpha, T^\beta, T^+, T^-]=\frac14
f^{\alpha\beta}_{\quad\gamma}T^\gamma+ l^{\alpha\beta}T^-+
g^{\alpha\beta} T^++k^{\alpha\beta}{}_A T^A%
\ee%
where $f^{\alpha\beta}_{\quad\gamma},\ l^{\alpha\beta},\
g^{\alpha\beta},\ k^{\alpha\beta}{}_A$ are some unknown coefficients
to be determined later. Taking anti-commutator of both sides of
\eqref{T+--closure} with $T^-$ we learn that coefficient of $T^+$ is
zero, $g^{\alpha\beta}=0$. Multiplying both sides with $T^+$ and
taking the trace (recall \eqref{T-2=0-four-bracket}) the left hand
side vanishes and therefore $l^{\alpha\beta}=0$.
 Commutator of
both sides of \eqref{T+--closure} with $T$, leads to
$k^{\alpha\beta}{}_A [T,T^A]=0$. On the other hand if we multiply
both sides of \eqref{T+--closure} with $T^-$ and then its commutator
with $T^+$  we learn that $k^{\alpha\beta}{}_A (2T^A+[T,T^A])=0$ and
hence $k^{\alpha\beta}{}_A T^A=0$.

Using the above and in particular $[T^\alpha,
T^\beta]=f^{\alpha\beta}{}_{\gamma}TT^\gamma$, that $TT^-=-T^-$
and that $Tr(T^\alpha T^-)=0$ one can show that trace of any
number of $T^\alpha$'s with $T^-$ is zero. This  in particular
implies that $f^{\alpha\beta\gamma}{}_{\lambda}h^{\lambda\rho}=0$.
$h^{\alpha\beta}$ is non-degenerate and invertible therefore,%
\be\label{f-4-index}%
f^{\alpha\beta\gamma}{}_{\lambda}=0\ , %
\ee%
and \eqref{f-k} reduces to $k^{\alpha\beta\gamma}{}_A  T^AT^-=0$
and moreover we have
\be\label{F-g}%
\{T^+, F^{\alpha\beta\gamma}T^-\}=-F^{\alpha\beta\gamma}\ T=
g^{\alpha\beta\gamma}\ \one + k^{\alpha\beta\gamma}{}_A \{T^+,
T^A\}\ .
\ee%

After the above analysis in summary we remain with
\bse\label{T-2=0-four-bracket-summary}%
\begin{align}%
[T^\alpha, T^\beta, T^\gamma, T^-]&= F^{\alpha\beta\gamma} T^-= g^{\alpha\beta\gamma} T^-+ k^{\alpha\beta\gamma}{}_AT^A\\
[T^\alpha, T^\beta, T^+, T^-] &= \frac14\,[T^\alpha, T^\beta] T=
f^{\alpha\beta}{}_\gamma T^\gamma \ ,
\end{align}%
\ese%
Furthermore, using \eqref{push-thru-trace} we learn that%
\be\label{f-g}%
f^{\alpha\beta\gamma}=f^{\alpha\beta}{}_\rho h^{\rho\gamma}=
-\frac12 Tr (\one)g^{\alpha\beta\gamma}\  .%
\ee%

To complete our analysis and to determine the yet unknown
coefficients $k^{\alpha\beta\gamma}{}_A$ and
$f^{\alpha\beta}_{\quad\gamma}$ we examine the relaxed fundamental
identity.
Let us first rewrite the identity for generic generators $T^a$:%
\be\label{fund-iden-Ta}%
\begin{split}%
&[[T^a,T^b,T^c,T^-]_{phys},T^d,T^e,T^-]_{phys}+[T^c,[T^a,T^b,T^d,T^-]_{phys},T^e,T^-]_{phys}+\cr
&+[T^c,T^d,[T^a,T^b,T^e,T^-]_{phys},T^-]_{phys}=
[T^a,T^b,[T^c,T^d,T^e,T^-]_{phys},T^-]_{phys}, %
\end{split}%
\ee%
where $T^a=T^\alpha, T^-$ or $T^+$. For three choices of the
$(abcde)$ indices the above fundamental identity does not trivially
hold, these cases are:\\
\emph{i)} $(abcde)=(\alpha\beta\gamma\rho +)$ implying%
\be\label{fg1-fund-iden}%
f^{\alpha\beta}_{\quad\sigma}g^{\gamma\rho\sigma}=f^{\gamma\rho}_{\quad
\sigma}g^{\alpha\beta\sigma}\ .%
\ee%
\emph{ii)} $(abcde)=(\alpha + \gamma\rho\lambda)$ implying%
\be\label{fg2-fund-iden}%
f^{\alpha\gamma}_{\quad\sigma}g^{\rho\lambda\sigma}+f^{\alpha\rho}_{\quad
\sigma}g^{\lambda\gamma\sigma}+f^{\alpha\lambda}_{\quad\sigma}g^{\gamma\rho\sigma}=0 .%
\ee%
\emph{iii)} $(abcde)=(\alpha + \gamma\rho+)$ implying%
\be\label{ff-fund-iden}%
f^{\alpha\gamma}_{\quad\sigma}f^{\rho\sigma}_{\quad\lambda}+f^{\rho\alpha}_{\quad
\sigma}f^{\gamma\sigma}_{\quad\lambda}+f^{\gamma\rho}_{\quad\sigma}f^{\alpha\sigma}_{\quad\lambda}=0 .%
\ee%
Recalling \eqref{f-g} the only independent of the above equations is
\eqref{ff-fund-iden}.

 Noting (\ref{T-2=0-four-bracket}b),
\eqref{T+--closure} and that
$[T,T^\alpha]=0$, it is seen that%
\be\label{TTalpha-algebra}%
 [TT^\alpha, TT^\beta]=f^{\alpha\beta}_{\quad\gamma}TT^\gamma,%
\ee%
therefore, recalling \eqref{ff-fund-iden}, $TT^\alpha$'s are
generators of a (classical) Lie-algebra which is a sub-algebra of
${\cal G}$, with the structure constants
$f^{\alpha\beta}_{\quad\gamma}$. We will denote this sub-algebra
by ${\cal H}$.

Given any classical Lie algebra ${\cal H}$, the only remaining
parameter in our brackets is $k^{\alpha\beta\gamma}{}_A$. As
discussed%
\be\label{kTA}%
k^{\alpha\beta\gamma}{}_A T^AT^-=0\ , %
\ee%
which can only be satisfied if either $k^{\alpha\beta\gamma}{}_A$
or $T^A T^-=0$. The first choice is not a possibility, because
there is no classical Lie-algebra other than $su(2)$ for which the
totally anti-symmetric three tensor%
\[F^{\alpha\beta\gamma}T =\frac{1}{12}\left(\{TT^\alpha, [TT^\beta,
TT^\gamma]\}+ \{TT^\gamma, [TT^\alpha, TT^\beta]\}+ \{TT^\beta,
[TT^\gamma, TT^\alpha]\}\right)\ ,
\]
is proportional to the identity. \footnote{We will return to the
special case of ${\cal H}=su(2)$ later in this section.} So, we
are forced to choose the
other possibility, i.e. %
\be\label{T-TA}%
T^A T^-= 0 \quad \Rightarrow \quad TT^A=-T^A T=-T^A. %
\ee%
We may solve the above   as%
\be\label{tildeTA}%
T^A=T^- {\tilde T}^A= {\tilde T}^A T^- \quad
\Leftarrow\!\Rightarrow \quad {\tilde T}^A=\{T^+, T^A\}\ ,
\ee%
where%
\be\label{tildeTA+-}%
[T^\pm, {\tilde T}^A]=[T, {\tilde T}^A]=0\ .%
\ee%
In terms of ${\tilde T}^A$, \eqref{F-g} is written as%
\be\label{F-tildeT}%
-F^{\alpha\beta\gamma}\ T= g^{\alpha\beta\gamma}\ \one +
k^{\alpha\beta\gamma}{}_A {\tilde T}^A\ . %
\ee%

To elaborate on the spurious sector and in particular the algebra of
the ${\tilde T}^A$'s, we examine the relaxed closure condition for the
brackets involving $T^A$. As it is seen from
(\ref{relaxed-closure-four-bracket}b) there are three such cases.
For brackets of the form $[T^a, T^b, T^A, T^-]$ when both $a$ and
$b$ are $\alpha$-type the bracket vanishes and the only
non-vanishing case is when
$(ba)=(+\alpha)$. After some algebra we find%
\be\label{+alphaA}%
12[T^{\alpha}, T^A, T^+, T^-]= [TT^\alpha, T^A]=C^{\alpha A}{}_{B}
T^B\ , %
\ee%
where the second equality is the statement of relaxed closure, with
some unknown constants $C$. From the above we also have%
\be\label{alphaA-com}%
[TT^\alpha, {\tilde T}^A]=C^{\alpha A}{}_B {\tilde T}^B\ . %
\ee%
Consistency of the above equation implies that%
\be\label{Cf}%
C^{\alpha A}{}_B C^{\beta B}{}_D-C^{\beta A}{}_B C^{\alpha
B}{}_D=f^{\alpha\beta}{}_\gamma C^{\gamma A}{}_D\ .%
\ee%
Since $T^AT^B=0$, brackets involving  two and three $T^A$'s
identically vanish. Given the above information we now proceed
to construct the underlying algebra ${\cal G}$.%\\
%\vskip .5mm%
 \paragraph{The algebra ${\cal I}$ constructed from $T, T^+$ and $T^-$ :}

As discussed above \eqref{T-def}, \eqref{T2}%
\[%
[T, T^-]=-2T^-, \quad  [T^+, T^-]=T, \quad \{T^+, T^-\}=\one .\]%
Thus, ${\cal I}$ will be identified once $[T, T^+]$ is known. It
is straightforward to show
\[
\{[T, T^+], T^-\}=2\cdot\one, \quad \{[T,T^+], T^\alpha\}=0, \quad
[[T,T^+], T^-]]=2T,\]%
and hence $[T^\alpha, T^\beta, [T, T^+], T^-]=12[T^\alpha,
T^\beta] T\in {\cal K}$. Therefore, $[T, T^+]$ is an element in
${\cal K}$, and since  its anti-commutator with
$T^-$ equals $2\cdot \one$, we conclude%
\be\label{TT+}%
[T, T^+]=+2 T^+.%
\ee%
This is also consistent with all other properties quoted above. As
a consequence, one can show that $(T^+)^2=0$.

To sum up, $T, T^+$ and $T^-$ form the following algebra%
\bse\label{cal-I}%
\begin{align}%
[T, T^\pm]=\pm 2 T^\pm, \quad  \quad &[T^+, T^-]=T\\
   \{T^+, T^-\}=\one \qquad & (T^-)^2=0.
\end{align}%
\ese%
Equations (\ref{cal-I}a) fix the algebra to be $su(2)$ while
(\ref{cal-I}b) fixes its representation to be $2\times 2$
matrices. An explicit solution to the above equations is%
\be\label{cal-I-exp-rep}%
T^-=\sigma^-= \frac12 (\sigma^1-i\sigma^2), \quad T=\sigma^3,\quad
T^+=\sigma^+=\frac12 (\sigma^1+i\sigma^2), %
\ee%
where $\sigma^i$ are the Pauli matrices. It is also noteworthy
that  $(T^+)^\dagger = T^-$, $T^\dagger=T$.

\paragraph{Fixing the underlying algebra ${\cal G}$ :}
The algebra ${\cal G}$ is obtained by studying the closure of the
commutators between the generators of its ${\cal I}$ and ${\cal
H}$ sub-algebras as well as the algebra constructed from ${\tilde
T}^A$'s, which will be denoted by ${\tilde{\cal H}}$. With the
above considerations (the commutator or) the algebra of ${\tilde
T}^A$'s will not be fixed. However, from \eqref{alphaA-com} and
\eqref{F-tildeT}  it is seen that ${\tilde{\cal H}}$ should
contain ${\cal H}$  as a subalgebra. Moreover, in general
${\tilde{\cal H}}$ may be taken as the {enveloping algebra} of
${\cal H}$, $Env({\cal H})$, or depending on ${\cal H}$, some
particular subalgebra of $Env({\cal H})$.

To complete our analysis it will be useful to give an explicit
representation for the underlying algebra ${\cal G}$. Based on
what we have discussed any element in ${\cal G}$, and in
particular $T^\alpha$, ${\tilde T}^A$, $T^\pm$ and $T$ can be
written as
\be\label{TTalpha}%
\begin{split}%
T^{\pm}=\one\otimes \sigma^{\pm}, &\qquad T=\one \otimes \sigma^3%
\cr%
 T^\alpha=t^\alpha \otimes \sigma^3, & \qquad
{\tilde T}^A={\tilde t}^A \otimes
\sigma^3\ , %
\end{split}\ee%
where $t^\alpha$ and ${\tilde t}^A$  are respectively generators
of ${\cal H}$ and ${\tilde{\cal H}}$, that is%
\be%\begin{split}%
[t^\alpha, t^\beta]=f^{\alpha\beta}_{\quad \gamma}\ t^\gamma\ ,
\qquad  [t^\alpha, {\tilde t}^A]=C^{\alpha A}_{\quad B}\ {\tilde
t}^B \ .
%\end{split}%
\ee%

As discussed neither of the algebras ${\cal H}$ and ${\tilde{\cal
H}}$ nor their representations are  fixed. However, as a general
solution one may take ${\tilde{\cal H}}=Env ({\cal H})$ (up to an
Abelian $u(1)$ factor) in which case, if we choose to work with
$N\times N$ representation of ${\cal H}$ the algebra ${\tilde{\cal
H}}$, independently of ${\cal H}$, will be $su(N)$ and therefore
${\cal G}=su(2N)$ (see, however, the comment below). A special case
which is physically well-motivated is ${{\cal H}}={\tilde{\cal H}}$.
For this case ${\cal H}$ is necessarily fixed to be $su(N)$ in
its \emph{fundamental} $N\times N$ representation. In this case \[
g^{\alpha\beta\gamma}=-\frac{1}{N} f^{\alpha\beta\gamma}\] where $f$
are the structure constants of $su(N)$.

%With the above one can see that%
%\be\label{TTalpha-com}%
%[T^\alpha, TT^\beta]=f^{\alpha\beta}_{\quad\gamma} T^\gamma,\qquad
%[T^\alpha, T^\beta]=f^{\alpha\beta}_{\quad\gamma} TT^\gamma, %
%\ee%
%and%
%\be\label{TpmTalpha}%
%\begin{split}%
%T^{\pm\alpha}\equiv  \frac12 [T^\pm, T^\alpha], \quad [T^\pm,
%TT^\alpha]=0, \quad [T, T^\alpha]=0, \quad %
%[T^{\pm\alpha}, T^{\pm\beta}]=0,\cr [T^{\pm\alpha},
%T^\beta]=T^\pm\{T^\alpha, T^\beta\}, \quad [T^{\pm\alpha},
%T^{\mp\beta}]=-\frac12\left([T^\alpha, T^\beta]\pm T\{T^\alpha,
%T^\beta\}\right)\ .
%\end{split}\ee%
%It is then readily seen that the set of $T^A=\{T^+, T^-, T,
%T^\alpha, TT^{\alpha}, T^{+\alpha}, T^{-\alpha}\}, \alpha=1,2,\cdots
%N^2-1.$ forms a complete basis for the $4N^2-1$ generators of the
%$su(2N)$ algebra. Explicitly, $T^A$ provides a $2N\times 2N$
%representation for the $su(2N)$:%
%\be\label{su(2N)-gen}%
%T^A\in \{ t^\alpha\otimes \one,\ \one\otimes \sigma^i,\
%t^\alpha\otimes \sigma^i,\}%
%\ee%
%where $t^\alpha$ are generators of $su(N)$ in $N\times N$ and
%$\sigma^i/2$ are generators of $su(2)$ in its $2\times 2$
%representation. Note that as is seen from \eqref{TTalpha-com}
%$T^\alpha$'s are not generators of $su(N)$, these are $TT^\alpha$
%which are $su(N)$ generators.

Before closing  this section, three comments are in order:
\begin{itemize}
\item For the very special case of ${\cal H}=su(2)$ and in its
fundamental $2\times 2$ representation, it is readily seen that
one can take $k^{\alpha\beta\gamma}{}_A$ to be zero. In this case
there is no need to introduce the spurious sector \cKs.
Nonetheless, for this case again the underlying algebra ${\cal G}$
will be $su(4)$.

\item Although we usually consider ${\cal H}$ to be a simple Lie-algebra,
it could also be a semi-simple algebra. The particular and
interesting example of this case is ${\cal H}=so(4)$. (Note,
however, that as discussed above this is not the Euclidean
three-algebra.) In this case, if we work with the $4\times 4$
representation of $so(4)$ algebra then we can take ${\tilde {\cal
H}}=so(4)\times u(1)\times u(1)$ in which case the two $u(1)$
factors are generated by $\gamma^5$ and the $4\times 4$ identity
matrix. With this choice the consistency relation \eqref{Cf} is
obviously satisfied. The underlying algebra ${\cal G}$ in this case
is $8\times 8$ representation of $su(4)\times su(4)$.

\item
 As we have discussed the underlying algebra in both of the
$(T^-)^2=\one/2$ and $(T^-)^2=0$ cases can be (and indeed for the
physically interesting ones is) an $su(2N)$ algebra. The $N=2$
case, related to the ${\cal G}=su(4)$, is very special because it
is isomorphic to $so(6)$. In this sense it may seem that for the
${\cal G}=su(4)$ case there are two different (Euclidean and
Lorentzian) solutions. But, it turns out that both of these
solutions are indeed physically the same and they are related by a
change of basis $T^a$'s: take ${\cal H}=su(2)$ and choose
$T^+-T^-=\one_2 \otimes i\sigma^2$ as $i{\gamma}^5$ and the four
$T^a$'s (the Dirac $\gamma$ matrices) to be  $2T^\alpha=
\sigma^\alpha\otimes \sigma^3$,  $\alpha=1,2,3$ and
$T^++T^-=\one_2 \otimes \sigma^1$. One should, however, note that
such a change of basis and taking the linear combination of
generators as new ``$T^-$'' does not generally work because the
fundamental identity is not linear in $T^-$. It is not difficult
to show, using direct examination of the fundamental identity,
that it only works for ${\cal H}=su(2)$. As a related comment, we
note that in the $(T^-)^2=\one/2$ case  $T^-$ is hermitian and in
the $(T^-)^2=0$ it is not. As we have discussed the $(T^-)^2=0$
case does not have a solution with hermitian $T^-$.
\end{itemize}

\section{The alternative representation for the BLG theory}

After replacing the BLG three-algebras with the
relaxed-three-algebras \crta\ and realizing the
relaxed-three-brackets with the four-brackets of usual matrices, we
are now ready to re-write the BL action in terms of usual matrices;
the only thing we need to do is to replace the three-brackets of the
BL action with the four-bracket and recall the definition of the
trace.
%For the two cases discussed in the previous section we will obtain two seemingly different actions.
As discussed we take our gauge fields to have $A_{i\ ab}$ and $A_{i\
aA}$ components and define the covariant derivative of any
element $\Phi$ in ${\cal K}\oplus {\cal K}_S$ as%
\be\label{cov-der-four-bracket}%
D_i\Phi=\partial_i\Phi- [T^a, T^b, \Phi, T^-]A_{i\ ab}-[T^a, T^A,
\Phi, T^-]A_{i\ aA}\ .%
\ee%
As shown the spurious parts of the field $\Phi$ as well as its
covariant derivative do not appear in the action (as they drop out
once we take the trace). Therefore, we can define a ``physical
gauge'' in which  $\Phi_A=0$ and $A_{i aA}$ components are chosen
such that \footnote{Note that due to the possibility of the presence
of $A_{i\ aA}$ components we have an extended notion of gauge
symmetry which allows for choosing these components of the gauge
fields. Since these components do not appear in the Chern-Simons
part of the action, this gauge symmetry is of course a trivial
symmetry of the corresponding BL action.} %
\be\label{physical-gauge}%
D_i\Phi= (D_i \Phi)_{phys}= \partial_i \Phi_{phys} - A_{i\ ab} [T^a,
T^b, \Phi, T^-]_{phys}=\left(\partial_i \Phi_d - f^{abc}{}_d A_{i\
ab}\Phi_c\right) T^d\ .%
\ee%

Equivalently, the ``physical gauge'' is the one in which $\Phi,\
D_i\Phi \in {\cal K}$. As discussed in the Lorentzian case, when we
choose $TT^\alpha$ to be Hermitian matrices, then $T^A$'s are not
Hermitian. Therefore in the physical gauge, when $T^A$ components are absent  we can demand Hermiticity%
\be\label{physical-gauge-2}%
\Phi^\dagger=\Phi\ , \qquad (D_i\Phi)^\dagger=D_i\Phi\ ,%
\ee%
where $\Phi$ are generic  scalar fields of the theory. In fact we
will be requiring the above conditions which also implies working
with non-spurious parts of fields.
 Hereafter we will always be working in the above mentioned
physical gauge \eqref{physical-gauge-2} and unless it is necessary
this point will not be mentioned explicitly. Therefore, in the
physical gauge the $A_{i\ aA}$ components do not appear and we only
remain with $A_{i\ ab}$ components of the gauge field.

Next we focus on the $A_{i\ ab}$ components. In general, $A_{ ab}$
can have $A_{-\alpha}$ and $A_{ \alpha\beta}$ components for the
$so(4)$-based algebra and $A_{\pm\mp}$, $A_{\pm\alpha}$ and
$A_{\alpha\beta}$ components for the Lorentzian algebras. However,
as it is seen from the explicit form of the covariant derivative
\eqref{cov-der-four-bracket} and also the form of the twisted
Chern-Simons action \eqref{twisted-CS}, not all of the possible
components of the gauge field appear in the action. For the
$so(4)$ based algebra it is only the $\alpha\beta$ component
\cite{BL1, BL2}, and for the Lorentzian case they are the
$+\alpha$ and $\alpha\beta$ components \cite{Verlinde, Russo}.
With the above definition, hence the other components, i.e.
$A_{-\alpha}$ for the $so(4)$-based case and $A_{\pm\mp}$ and
$A_{-\alpha}$ for the Lorentzian case, are ``gauge degrees of
freedom'' and may be chosen freely and for example can be set to
zero. It is also seen that the $T^-$ component of the $\Phi\in
{\cal K}$, for both the Euclidean and Lorentzian cases, is also a
free field not interacting with the other components.

\subsection{Lagrangian in terms of Four-brackets}

{}From the discussions of previous section and our construction of
three-brackets and the relaxed-three-algebras it is evident that
if in the action \eqref{BL-action} we replace three-brackets with
our prescribed four-brackets we will obtain a supersymmetric and
gauge invariant action. For both cases, Euclidean and Lorentzian,
the supersymmetry transformations and Lagrangian are alike. For
completeness we only show the explicit form of the action, its
equations of motion and supersymmetry and gauge transformations.
\paragraph{The action}
\be\label{MBL-action-four-bracket} \begin{split}%
  S&=\int d^3\sigma\ Tr\biggl[-\half D_i
  X^ID^i X^I-\frac{1}{2.3!}[X^I,X^J,X^K,T^-][X^I,X^J,X^K,T^-]\cr
 &\qquad+\frac{i}{2}\bar{\Psi}\gamma^i
  D_i\Psi-\frac{i}{4}[\bar{\Psi},X^I,X^J,T^-] \Gamma^{IJ}\Psi\cr%
   &\qquad+\frac{1}{2}\epsilon^{ijk}\left(A_{i\ ab}\partial_j A_{k\
cd} T^d+\frac{2}{3} A_{i\ ab} A_{j\ de} A_{k\ cf} [T^d, T^e, T^f, T^-]\right) [T^a, T^b, T^c, T^-]
\biggr].%
\end{split}\ee %
\paragraph{Equations of motion}%
\be\label{eom-four-bracket}\begin{split} %
 &\left(\gamma^i D_i\Psi+\half\Gamma^{IJ}[X^I,X^J,\Psi,T^-]\right)_{phys}=0\cr
 &\left(D^2X^I-\frac{i}{2}\Gamma^{IJ}[\bar{\Psi},X^J,\Psi,T^-]+\half[X^I,X^J,[X^I,X^J,X^K,T^-],T^-]\right)_{phys}=0\cr
 &\left(\tilde{F}_{ij}^{ab}+\epsilon_{ijk}\big(D^k X^I[T^a,T^b,X^I,T^-]
 -\frac{i}{2}\bar{\Psi}\gamma^k[T^a,T^b,\Psi,T^-]\big)\right)_{phys}=0
\end{split}\ee %
where $\tilde{F}$ is appeared in \eqref{e.o.m}.
\paragraph{Supersymmetry transformations}%
\be\label{SUSY-trans-four-bracket}\begin{split} %
 \delta X^I&=i\bar{\epsilon}\Gamma^I\Psi \cr %
 \delta\Psi&=D_i X^I\Gamma^I\gamma^i\epsilon-\frac{1}{6}[X^I,X^J,X^K,T^-]\Gamma^{IJK}\epsilon \cr%
 \delta (D_i\Phi)- D_i (\delta\Phi)
 &=i\bar{\epsilon}\gamma^i\Gamma^I[X^I,\Psi,\Phi,T^-],\ \ \ \  \forall \Phi
\end{split}\ee %
where it is understood that we are only considering the physical
parts of the fields. It is immediate to see that the action is
invariant when we also include non-physical and spurious parts in
the above supersymmetry transformations. Nonetheless, along the line
of arguments of \cite{BL2} one can show that the supersymmetry
algebra (i.e. commutator of two successive supersymmetry
transformations) does not close to a translation, up to gauge
transformations.

\noindent\textbf{Gauge transformations}%

We should emphasize that the following ``gauge transformations''
are the gauge symmetry remaining after fixing the physical gauge
\eqref{physical-gauge} and \eqref{physical-gauge-2}.

\emph{The Euclidean case}%
\be\begin{split}\label{gauge-trans-so4}%
\delta \Phi_ a&=\epsilon^{cdb}_{\quad\ a}\Lambda_{cd} \Phi_b\ ,\qquad \delta\Phi_-=0\\
\delta A_{i\ ab}&=\partial_i \Lambda_{ab}- \epsilon^{dec}{}_{
[a}\Lambda_{b] c} A_{i\ de},\qquad a,b,c,d=1,2,3,4\ .
\end{split}\ee %

\emph{The Lorentzian case}%
\be\begin{split}\label{gauge-trans-su2N-Phi}%
 \delta\Phi_{\alpha}&=f^{\beta\gamma}_{\ \ \ \alpha}(2\Lambda_{+\beta}\Phi_\gamma
 +\Lambda_{\beta\gamma}\Phi_+)\cr
 \delta \Phi_+&=f^{\alpha\beta\gamma}\Lambda_{\alpha\beta}\Phi_\gamma\cr %
 \delta\Phi_-&=0
\end{split}\ee %
\be\begin{split}\label{gauge-trans-su2N-Amu}%
\delta A_{i\ +\alpha}&=\partial_i \Lambda_{+\alpha}+2f^{\beta\gamma}_{\quad \alpha}\Lambda_{+\gamma}
A_{i\ +\beta}\\
\delta A_{i\ \alpha\beta}&=\partial_i \Lambda_{\alpha\beta}-2
f^{\rho\gamma}{}_{ [\alpha}\ \Lambda_{\beta]\gamma}\ A_{i\
+\rho}-f^{\rho\gamma}{}_{[\alpha} \Lambda_{\beta]+}\  A_{i\
\rho\gamma}
\end{split}\ee %
In the Lorentzian case the Greek indices are ranging from
$1,\cdots, dim {\cal H}$ and correspond to ${\cal H}$ indices.

\subsection{On the physical interpretation of the Lorentzian case}

As has been discussed in the literature  the $so(4)$-based
theories describe (the low energy limit of) two M2-branes on an
orbifold \cite{Lambert-Tong, Distler:2008mk}. The physical
interpretation of the Lorenztian case, however, is less clear. In
the \emph{usual treatment} all the components of the scalars
$X^I$, including $X^+$ and $X^-$ are taken to be real and hence
the negative signature in the metric $h_{ab}$ means that one
combination of $X^+$ and $X^-$ has negative eigenvalue, in other
words, we have ghosts. Existence of ghosts which couple to the
other fields endangers the unitarity of the theory. \emph{Our
treatment} of the three-algebras, however,  sheds light on the
unitarity or ghost problem of the Lorentzian case.

As shown in section 3, the negative eigenvalue of the metric is
indeed a reflection of the way we realize the three-brackets and
the way $T^\pm$ are embedded in the underlying ${\cal G}$ algebra.
Therefore, in contrast to the usual treatment in our description,
while the scalar field $X^I=X^I_{\ a} T^a$ is still Hermitian
$X^+$ and $X^-$ are not,
explicitly%
\be\label{X-Hermiticity}%
(X^I)^\dagger= X^I \ \  \Rightarrow\ (X^I_+)^*=X^I_-\ . %
\ee%
With the above it is immediate to see that we do not have the
negative kinetic term, or ghost problem. Nonetheless, the unitarity
problem shows up in some other place: the interaction terms in the
Hamiltonian only involve $X^I_+$ (and not $X^I_-$) and hence the
Hamiltonian in our description is not Hermitian.

To resolve the problem  we recall the gauge symmetry of our
Lagrangian and the fact that $X^I_+$ components are \emph{not}
gauge invariant \eqref{gauge-trans-su2N-Phi} and hence are not
directly physical observables. This opens up the possibility that
this non-Hermiticity can be an artifact of the gauge symmetry and
the physical theory is indeed Hermitian and unitary. In what
follows we argue that there is a gauge, the \emph{Hermitian
gauge}, in which the Hamiltonian is explicitly Hermitian,
resolving the problem with
unitarity. %
\subsubsection{The Hermitian gauge}

As is seen from \eqref{gauge-trans-su2N-Phi} the gauge
transformations are parameterized through two sets of gauge
parameters $\Lambda_{+\alpha}$ and $\Lambda_{\alpha\beta}$, each
having $dim {\cal H}$ number of parameters. Moreover, $X^I_+$ only
transforms under the $\Lambda_{\alpha\beta}$-type gauge
transformations while is invariant under the
$\Lambda_{+\alpha}$-type transformations.

On the other hand, the Hamiltonian becomes Hermitian only if $X^I_+$
and $X^I_-$ are equal up to a sign, that is when $X^I_+$ is real or
pure imaginary. Therefore, if we fix the $\Lambda_{\alpha\beta}$-gauge such that%
\[
 X^I_+=\pm X^I_-,
\] %
the Hamiltonian becomes Hermitian. To fix the sign choice in the
above gauge fixing expression we choose the gauge such that the
positivity of the Hamiltonian (the potential) is ensured. It is
straightforward to check that this is fulfilled with the negative
sign. The appropriate \emph{Hermitian-gauge} fixing condition is
then \footnote{To fix the gauge condition \eqref{Hermitian-gauge}
we in fact need at least eight gauge parameters. Therefore, our
arguments works for $dim {\cal H}\geq 8$. As will become clear in
the next subsection the appropriate ${\cal H}$ for $N$ M2-branes
is $su(N)$, this corresponds to $N\geq 3$. For the special case of
$N=2$, which as discussed in the end of section 4 is equivalent to
the $so(4)$-based algebras with an appropriate change of basis,
one can explicitly show that in this specific gauge the two
Lorenztian and Euclidean descriptions are indeed identical, of
course once an $su(2)$ part of the $so(4)$
gauge symmetry of the latter case is also fixed.}%
\be\label{Hermitian-gauge}%
X^I_+\ + X^I_ -=0\ . %
\ee%
One should note that the above gauge fixing condition only
partially fixes the $\Lambda_{\alpha\beta}$ gauge
symmetry.\footnote{Noting the comments in footnote 5, only for
$N=3$ these gauge transformations can be completely fixed by
\eqref{Hermitian-gauge}.} Besides the Hermiticity problem of the
Hamiltonian, the above gauge also removes half of the degrees of
freedom in $X^I_\pm$. Hereafter,
we will work in the \emph{Hermitian gauge} and define%
\be\label{YI-c.o.m}%
Y^I\equiv -\frac{i}{2}(X^I_+ - X^I_-)=-iX^I_+\ . %
\ee%

After fixing the $\Lambda_{\alpha\beta}$-type gauge transformations,
we only remain with $\Lambda_{+\alpha}$. For this restricted gauge
symmetry the gauge transformations are%
\be\begin{split}%
 \delta\Phi_{\alpha} &=2f^{\beta\gamma}_{\ \ \
 \alpha}\Lambda_{+\beta}\Phi_\gamma\\
 %=[\Lambda,\Phi]_\alpha \\
\delta A_{i\ +\alpha}&=\partial_i
\Lambda_{+\alpha}+2f^{\beta\gamma}_{\quad \alpha}\Lambda_{+\gamma}\
A_{i\ +\beta}\\
\delta A_{i\ \alpha\beta}&=
%\partial_i \Lambda_{\alpha\beta}-2 f^{\rho\gamma}_{\quad [\alpha} \Lambda_{\beta]\gamma} A_{i\ +\rho}
-f^{\rho\gamma}{}_{ [\alpha} \Lambda_{\beta]+}\  A_{i\
\rho\gamma}.
\end{split}\ee %
After the following renaming
\be\begin{split}%
\hat\Lambda_{\alpha} =\frac12\Lambda_{+\alpha}\ ,\quad {\hat
A}_{i\ \alpha} =\frac12 A_{i\ +\alpha}\ ,\quad {\hat B}_{i\
\gamma} = f^{\alpha\beta}_{\ \ \gamma} A_{i\
\alpha\beta},%
\end{split}\ee%
the above gauge transformations take the familiar form of
standard gauge transformations for the algebra ${\cal H}$ with
$\hat A_i$ as the gauge field and the two ``matter fields"
$\Phi_\alpha$ and ${\hat B}_{i\ \alpha}$ in the adjoint (and
anti-adjoint) representations:
\be\label{gauge-trans-hatted}%
\begin{split}%
\delta\Phi&=[\hat\Lambda, \Phi]\ ,\\
\delta {\hat B}_i &=-[\hat\Lambda, {\hat B}_i]\ , \\
\delta {\hat A}_i &= {\hat D}_i \hat\Lambda=\partial_i \hat\Lambda
-[\hat\Lambda, A_i]
\end{split}\ee%
where%
\[ ([\hat\Lambda, \Phi])_\alpha=f^{\beta\gamma}{}_{\alpha}
\hat\Lambda_\beta\Phi_\gamma.
\]
It is evident that $Y^I$ are singlets and does not transform under
the above gauge transformations of the ${\cal H}$ algebra. As we
see after fixing the \emph{Hermitian gauge} the proposed \DtNe
action \eqref{MBL-action-four-bracket} written in terms of hatted
fields and $Y^I$ \eqref{YI-c.o.m} exhibits a standard ${\cal H}$
invariance (with the gauge transformations
\eqref{gauge-trans-hatted}).

\subsection{Connection to multi M2-brane theory, the parity invariance}

The proposed BL \DtNe theory is expected to be related to theory
of multiple M2-branes in an eleven dimensional flat space
background. As such, one then expects that this theory should have
the same form for a system of M2-branes and anti-M2 branes. From
the worldvolume theory viewpoint M2-branes and anti M2-branes are
related by the worldvolume parity and hence the proposed BL theory
should be parity invariant \cite{Schwarz, Verlinde, BL-comment}.
In terms of our four-bracket and the algebra ${\cal G}$, the
parity invariance is respected if the parity is defined as%
\be\label{parity-def}%
\sigma^0,\sigma^1\to \sigma^0,\sigma^1, \quad\sigma^2\to -\sigma^2,
\quad T^\pm\to -T^\pm, \quad T^\alpha\to
T^\alpha,%
\ee%
($\sigma^0,\sigma^1$ and $\sigma^2$ are M2-brane worldvolume
coordinates) while $X^I$ behave as scalars under parity, $\Psi$ as
a $3d$ fermion, and $A_\mu$ as a $3d$ vector. That is, under parity%
\be\label{parity-on-fields}%
\begin{split}%
X^I_\alpha\to X^I_\alpha\ ,& \quad X^I_\pm\to -X^I_\pm\ ,\\
(A_0, A_1, A_2)_{\ \alpha\beta}\to (A_0, A_1, -A_2)_{\
\alpha\beta}\ ,  &\quad (A_0, A_1, A_2)_{\ +\alpha}\to (-A_0,
-A_1, A_2)_{\ +\alpha}\ .
\end{split}\ee%
As we see the parity \eqref{parity-def} is an automorphism on the
algebra $\cal G$ as well as its subset $\cal K$ over which the
four-bracket closes (in the relaxed closure sense). More
precisely, under the above parity the ${\cal H}\in {\cal G}$ is
invariant, while on the $su(2)\in {\cal G}$ it acts as an
automorphism.

It is also immediate to check that with \eqref{parity-on-fields}
the action \eqref{MBL-action-four-bracket} is parity invariant.
Moreover, the \emph{Hermitian gauge} \eqref{Hermitian-gauge} is
preserved under parity. This is a necessary condition to have a
consistent (Hermitian) multi M2-brane theory.

So far, for the Lorentzian case we have not identified the algebra
${\cal H}$ and the underlying algebra ${\cal G}$. In the next
section we will argue that the choice ${\cal H}=su(N),\ {\cal
G}=su(2N)$ corresponds to the low energy limit of $N$ M2-branes.

\subsection{Analysis of 1/2 BPS states }

To relate the above ``gauge fixed relaxed BLG model'' to the
theory of multiple M2-branes, we need to specify the algebra
${\cal H}$ and relate that to the number of M2-branes $N$. This
can be done by studying the half BPS configurations of the model,
the moduli space of which should be identified with the moduli
space of $N$ membranes in eleven dimensional flat background,
which is $\mathbb{R}^{8N}/S_N$.

The half BPS sector is the one for which the right-hand-side of
supersymmetry transformations \eqref{SUSY-trans-four-bracket}
vanishes for any arbitrary supersymmetry transformation parameter
$\epsilon$. In order $\delta X^I$ and $\delta A_{i ab}$ to vanish
we need to turn off the fermionic field $\Psi$. We are then left
with the fermionic transformation which has two terms. These terms
come with different matrix structure in $so(2,1)$ and $so(8)$
gamma-matrices. Therefore, for $\delta \Psi$ to vanish for
any $\epsilon$ each term should vanish independently, i.e.%
\bse\label{1/2BPS}%
\begin{align}%
D_i X^I &=0 \ ,\\
[X^I, X^J, X^K, T^-]_{phys} &= 0.%
\end{align}%
\ese%
Recalling the equations of motion \eqref{eom-four-bracket},
demanding vanishing of (\ref{1/2BPS}a), the field strength of the
gauge field vanishes and one can always work in a gauge in which
$A_i=0$, and hence \eqref{eom-four-bracket} implies that
$\partial_i X^I=0$ In other words 1/2 BPS membranes must be flat
membranes with worldvolume $\mathbb{R}^{2,1}$. We are then left
with (\ref{1/2BPS}b) which recalling the definition of the
four-bracket, is satisfied if and only if%
\be\label{XXX}%
[X^I, X^J]=0\ .%
\ee%
Note that since we are working in the ``physical Hermitian gauge''
$X^I$ in the above equation have components only along the
$T^\alpha$ directions. Therefore, \eqref{XXX} is only satisfied
when $X^I$ are in Cartan subalgebra of ${\cal H}$ and that the
number of such possible $X^I$ matrices is $rank ({\cal H})-1$.
(Note that we have already taken out the ``center of mass'' degree
of freedom in $X_+^I$.) Noting that $X^I$ are basically related to
the position of M2-branes, this means that number of M2-branes $N$
minus one is to be taken as rank of ${\cal H}$.

As discussed in \cite{Russo,Verlinde} (see also \cite{HIM})
another test for the theory of multi M2-branes is that upon
``compactification'' it should reproduce theory of multi
D2-branes. This together with the above discussions fixes ${\cal
H}={\tilde{\cal H}}=su(N)$ in its fundamental $N\times N$
representation as the theory of $N$ membranes and hence the
underlying algebra ${\cal G}=su(2N)$. With this choice it is
evident that the moduli space of solutions to \eqref{XXX} is the
desired $R^{8N}/S_N$.

Let us discuss some low-lying $N$'s in more detail. The $N=1$
corresponds to a single M2-brane. In this case the fields are
$2\times 2$ matrices and therefore all the four-brackets vanish.
In this case, as expected, we are dealing with a non-interacting
free theory and the only remaining degree of freedom are $Y^I$
(and their fermionic counterparts). This is suggesting that $Y^I$
should correspond to the center of mass degree of freedom in the
$N>1$ cases.\footnote{It is worth noting that under parity $Y^I\to
-Y^I $ and hence the parity transformation we have introduced here
besides changing an M2-brane to an anti M2, also acts as a parity
on the target space directions transverse to the brane. In the
static gauge for the M2-brane, this means that under our parity we
are essentially changing sign on nine space coordinates of the
eleven dimensional background. This transformation is also a
symmetry of the eleven dimensional supergravity and expected to be
symmetry of the M2-brane theory too.}

The next case is $N=2$ corresponding to two M2-brane system which
has $X^I_\alpha$ fields in the adjoint of $su(2)$ plus the $Y^I$'s
which are $su(2)$ singlets. Here we are dealing with $4\times 4$
representation of $su(4)$. As discussed this case also makes
connection with the $so(4)$-based algebras which have also been
discussed to correspond to the two M2-brane dynamics. To argue for
the claim that $Y^I$ are the center of mass coordinates one should
show that they decouple from the dynamics. The first steps toward
this end has been taken in \cite{Sen1, Sen2}, further arguments in
support of this is postponed to future works \cite{progress}. Once
this claim is established for $N=2$, the same argument can then be
generalized to a generic $N$.

\section{Discussion and Outlook}

In this work we have studied the structure of three-algebras which
appear in connection with formulation of \DtNe\ theories. These
algebras are also thought to be relevant to the quantization of
Nambu three-brackets and hence quantized (multi) M2-brane
theories. We explored formulation of the idea that the
non-associative three-algebras and their representations can be
expressed in terms of inherently associative classical
Lie-algebras (and their matrix representations), by introducing
the ``non-associative bracket structure'' on these algebras; we
denoted the underlying associative matrix algebra by ${\cal G}$.
We argued that to keep the essential properties of the
non-associative three-brackets, when expressed in terms of
matrices, we need to replace the three-bracket with a four-bracket
which is defined as the totally anti-symmetric product of matrices
appearing in the bracket. In this procedure, we then need to
introduce a given extra matrix, which was called $T^-$, when
moving from a three-bracket to a four-bracket. ($T^-$ is of course
an element in \cG.)

With the working assumption that $T^-$ should anti-commute with
all the elements of the ``three-algebra'', we examined the
necessary closure and fundamental identity. As argued, however,
one can still have the notion of physically interesting
three-algebras if we relax both the fundamental identity and
closure conditions in a very particular way. This was done by
demanding the closure of the brackets up to the \emph{spurious}
parts of the elements of the algebra. In other words, any element
has a \emph{physical} as well as a \emph{spurious} part and only
bracket of physical parts of the elements lead to a physical
element, and the physical part of brackets satisfy the fundamental
identity. With this extended, generalized or relaxed notion of
fundamental identity and the closure we hence defined the
\emph{relaxed-three-algebras} \crta .

As we showed the above definition of \rta s is still restrictive
enough to fix the possible underlying algebra ${\cal G}$ and its
representations. We showed that within our working assumptions only
two cases are possible, one corresponding to the case with positive
definite metric on the \rta, the Euclidean case, and the other with
a Lorentzian metric on that algebra. We should emphasize that in our
analysis we did not assume anything about the signature of the
metric and this condition appeared as the consistency condition
within our setting. Moreover, as discussed there is nothing
inherently Lorentzian in the underlying algebra ${\cal G}$ and the
Lorentzian signature is as an artifact of the choice of the set of
generators of ${\cal G}$ which appear in the four-brackets. This is
the resolution to the problem of  the negative kinetic energy states
(ghosts) in the usual treatment of the Lorentzian BL theory \cite{
Sen1, Sen2, ghost-gauging}. For the Euclidean case, using the
results of \cite{Papadopoulos:2008sk}, we concluded that there is
only one possibility which was called the ``$so(4)$-based'' algebras
for which the underlying algebra is $su(4)$. For the Lorenzian case,
however, we showed that there remains a freedom in choosing the
algebra which was then fixed once the setting of \rta s was employed
in the multi M2-brane theory. As argued the Euclidean case can be
formulated without the spurious parts for elements, whereas spurious
parts are necessary for the Lorentzian case.

In the corresponding physical model the spurious parts of the fields
do not appear at all and the Hilbert space of physical states is
hence defined by modding out the total Hilbert space by the spurious
parts. As discussed in the specific physical model of multi
M2-branes the spurious parts are reminiscent of usually overlooked
gauge symmetries. This spurious parts are very similar to the same
concept in the context of $2d$ $CFT$'s and in string theory
\cite{Polch1}. Exploring and understanding these symmetries seems to
be an important clue to better understanding of, and resolution to,
one of the fundamental open issues in the Bagger-Lambert multi
M2-brane theory for more than two M2-branes.

Analyzing the moduli space of 1/2 BPS states of the new
realization of the BL-theory in terms of four-brackets, we argued
that in order this moduli space to be the same as what is expected
from $N$ M2-branes in flat 11 dimensional background, the
underlying algebra ${\cal G}$ must be taken  $su(2N)$ and the
physical fields and states must be labeled by physical $N\times N$
representation of $su(N)$.

It is desirable to find a more intuitive picture for the role of the
$su(2N)$ underlying algebra and the $su(N)$ ${\cal H}$ algebra, as
we have for multiple (coincident) D-brane where the degrees of
freedom corresponding to open string attached to and stretched
between parallel D-branes leads to the $su(N)$ structure
\cite{Witten}. Note that to get the $su(N)$ structure we should
remember that open strings stretched between D-branes come in two
opposite orientations  each of which includes a massless (vector)
state when two D-branes become coincident. For the case of
M2-branes, similar to the D-brane case, we have open M2-branes
stretched between two M2-branes. Although we do not know spectrum of
open M2-branes as well as we want to, it is expected  that there are
massless states in the coincident M2-brane limit. Again similarly to
the stretched open string case, there are open M2 and anti-M2
branes. Recalling that M2-branes are two dimensional (to be compared
with one dimensional strings), for the case of membranes
 there are two options to get an anti-M2 brane for a given
M2; the M2-brane and anti-M2-brane are related by parity on the
worldvolume of the brane. This is suggestive that when we consider
the four possible open M2 and anti-M2 branes (two M2-brane in which
orientation on both directions have changed with respect to each
other and two respective anti-M2-branes which are related by
worldvolume parity to the two M2-brane cases) we are over-counting
the degrees of freedom and this should be mod out by the worldvolume
parity. In other words, the reduction from $su(2N)$ to $su(N)$ which
labels $N$ M2-brane fluctuations could be done through worldvolume
parity. As argued the parity on the M2-brane worldvolume is acting
as an automorphism of this $su(2N)$ algebra and keeps the $su(N)$
labels of the physical states/fields invariant. It would be very
interesting to make the above picture more precise and concrete
\cite{progress}.

As discussed the ghost problem of the Lorentzian three-algebras in
our setup manifested itself in our setup as non-hermiticty of the
Hamiltonian before the gauge fixing and can be removed once we fix
the \emph{Hermitian gauge}. This resolution which is in accord with
proposal in \cite{Sen1, Sen2}, however, requires identifying the
mode, which we called $Y^I$ as the center of mass degree of freedom
of M2-brane system. The problem which is still remaining in this
direction is establishing the fact that the center of mass degree of
freedom is indeed decoupled. Once this problem is settled, our setup
which is based on usual matrices provides the needed tools to make
further analysis of the \DtNe or  the multi-M2-brane theory.

%*******************************************************************************************************************************************
\section*{Acknowledgements}

M.M.Sh-J. would like to thank Sadik Deger and Henning Samtleben  for
fruitful discussions. J.S. would like to thank the IPM in Tehran for
hospitality during the early stages of this project.

%**************************************************************************************************************************

\end{document}